\documentclass[final,3p,times]{elsarticle}

\usepackage{graphicx}

\usepackage{amssymb}
\usepackage{amsmath}	% без этого не работаю ссылки на уравнения \eqref
\newcommand{\sech}{{\rm sech}}	% hyperbolic secant

\biboptions{compress}

\journal{arXiv.org}

\begin{document}

\begin{frontmatter}



\title{Excitation of high-amplitude localized nonlinear waves as a result of interaction of kink with attractive impurity in sine-Gordon equation}


\author[bsu]{E. G. Ekomasov}
\ead{EkomasovEG@gmail.com}
\ead{http://soliton.fizfaka.net/}

\author[bsu]{A. M. Gumerov}
\ead{bgu@bk.ru}

\author[bsu]{R. R. Murtazin}
%\ead{MurtazinRR@mail.ru}

\author[bsu]{A. E. Ekomasov}
%\ead{fean0r0@mail.ru}

\author[ipsm]{S. V. Dmitriev}
\ead{dmitriev.sergey.v@gmail.com}

\address[bsu]{Bashkir State University, Russia, 450076, Ufa, Z.Validi Str., 32}
\address[ipsm]{Institute for Metals Superplasticity Problems, Russian Academy of Science, Russia, 450001, Ufa, Khalturin Str. 39}

\begin{abstract}
We study properties of the localized solitons to the sine-Gordon equation excited on the attractive impurity by a moving kink. The cases of one- and two-dimensional spatially extended impurities are considered. For the case of one-dimensional impurity the possibility of excitation of the first even and odd high-amplitude impurity modes by the moving kink is demonstrated. For the case of two-dimensional impurity we show the possibility of excitation of the nonlinear high-amplitude waves of new type called here breathing pulson and breathing 2D soliton. We suggest different analytical expressions to model these nonlinear excitations. The dependencies of the oscillation frequency and the amplitude of the excited impurity modes on the impurity parameters are reported.
\end{abstract}


\end{frontmatter}

% \linenumbers

\section{Introduction}
\label{Introduction}

In the last years, the solitary wave dynamics has attracted
increasing attention of researchers \cite{Christiansen2000}. This
is due to the fact that solitons initially studied in integrable
systems gave rise to the study on solitary waves in non-integrable
systems to describe a number of physical problems. For example,
the sin-Gordon equation (SGE) is used for modeling wave
propagation in geological media, in molecular biology, field
theory models, in elementary particle physics, to name a few
\cite{Scott2004,Yakushevich2002}. The SGE soliton solutions help
to describe domain walls in magnetics, dislocations in crystals,
fluxons in Josephson's junctions, etc.
\cite{PMM2008eng,Kivshar2008eng,Gulevich2006}. In many cases
behavior of solitons can be effectively described by
quasi-particles and then their dynamics can be presented by
ordinary differential equations \cite{Fogel1977}. However, in the
presence of perturbations, the structure of solitons can be
changed and they can be more accurately described by deformable
quasi-particles \cite{Kivshar2008eng}. Soliton's internal degrees
of freedom can be excited in this case and they can play a very
important role in a number of physical processes
\cite{Kivshar1998,Kivshar1991a}. Soliton internal modes can be
responcible for the non-trivial effects of their interactions
\cite{D1,D2,D3,D4,D5}. The internal modes can include the
translational and pulsation mode describing long-lived
oscillations of soliton width
 \cite{Gonzales2002}. Effect of various perturbations on the excitation of
SGE soliton's internal modes attracts a lot of attention of
researchers. Local inhomogeneities are ubiquitous in many physical
systems, including those described by SGE, and it is very
important to study the soliton scattering on such impurities
 \cite{Kivshar2008eng}. For instance, there exist many works devoted to the
analysis of external force that varies in time and space
\cite{Kivshar2008eng,Kivshar1991a,Gonzales2002,Chacon2008,Willis2006_to_Quintero2000}.
Weak perturbations on the SGE solutions can be studied in frame of
the perturbation theory well-developed for solitons
\cite{Kivshar2008eng,Fogel1977}  but the effect of strong
perturbations can be analyzed only numerically
\cite{Bratsos2007,Currie1977,PMM2006}.

The effect of spatial modulation (inhomogeneity) of the periodic
potential or the presence of an impurity in the system are also
very interesting  \cite{Kivshar2008eng}. Depending on the geometry of the
system, physically meaningful can be one-dimensional or
multi-dimensional problems. The problem of scattering of SGE kinks
on impurities in one-dimensional case has been under consideration
for a long time  \cite{Fogel1977,Currie1977,Paul1979,Zhang1992}. For example, the model of
classical particle is applicable to the problem of kink-impurity
interaction in the case when impurity itself does not support
vibrational modes localized on the impurity  \cite{Kivshar2008eng}. Importance
of the impurity modes in the mechanisms of kink-impurity
interactions has been demonstrated in the works
 \cite{Kivshar2008eng,Piette2007kinks,Piette2005,Piette2007soliton,Javidan2008}. Let us mention such an interesting effect as
the reflection of kink by an attractive impurity due to the
resonance energy exchange between translational kink's mode and
the impurity mode.

Two-dimensional SGE has been also studied for a long time with the
use of analytical methods  \cite{Dodd1982,Qing2010,Koutvitsky2005} and with the help of
numerical methods  \cite{Bratsos2007,Dodd1982,Dehghan2008}. For example, in the works
 \cite{Bratsos2007,PMM2009eng} the appearance and motion of flexural solitary waves
on the kink interacting with a two-dimensional impurity has been
investigated. However, the possibility of excitation of various
two-dimensional vibrational modes localized on the impurity was
not discussed in those works.

In the present study we analyze the interaction of SGE kink with
impurity in the case when large-amplitude nonlinear waves
localized on the impurity are excited bas a result of interaction.

Let us consider the system defined by the following Lagrangian
\begin{eqnarray}
    \label{Lagrangian}
    L(x,y,t) = \frac{1}{2}
            \left[
                \left( \frac{\partial\theta}{\partial t}\right)^2
                + \left( \frac{\partial\theta}{\partial x}\right)^2
                + \left( \frac{\partial\theta}{\partial y}\right)^2
            \right]
            \pm \left[
                1 - \Delta K\left( x,y \right)
            \right]
            \sin^2 \theta .
\end{eqnarray}
The corresponding equation of motion for the scalar field
$\theta(x,y,t)$ has the following form
\begin{eqnarray}
    \label{EqMotion}
    \frac{\partial^2\theta}{\partial t^2} - \frac{\partial^2\theta}{\partial x^2}
    - \frac{\partial^2\theta}{\partial y^2}
    + \frac{1}{2}K\left( x,y \right)\sin 2\theta = 0,
\end{eqnarray}
where the function $K(x,y)$ defines the interaction of the field
$\theta(x,y,t)$ with the impurity.

\section{One-dimensional case}
\label{1Dcase}

The sine-Gordon model with the impurity extended in one dimension
is defined by the Lagrangian (\ref{Lagrangian}) with
\begin{eqnarray}
    \label{Impurity}
    K(x) = \left\{
                \begin{array}{ll}
                    1, & x<x_1,\,\,x>x_2+W \\
                    1-\Delta K, & x_1\le x \le x_2+W
                \end{array}
            \right.
\end{eqnarray}
where $W$ is the width of the impurity. Clearly, $\Delta K > 0$
describes a potential well, while $\Delta K < 0$ describes a
potential barrier.

In the case $K(x) = 1$ $(\Delta K=0)$ Eq. \ref{EqMotion}
supports the exact solution in the form of topological soliton or,
in other words, kink:
\begin{equation}
    \label{Kink}
    \theta(x,t) = 2 \arctan \left(
        \exp [ \gamma (v_0) (x-v_0 t) ]
    \right),
\end{equation}
where $\gamma (v_0)=(1-v_0^2)^{-1/2}$ with a parameter $0<v_0<1$
defining the velocity of the kink. Equation \eqref{EqMotion} also
admits the periodic in time solution in the form of breather,
\begin{equation}
    \label{Breather}
    \theta(x,t) = 2 \arctan \left(
        \frac{
            \sqrt{1-\omega^2} \sin \omega t
        }{
            \omega \cosh \left[ \sqrt{1-\omega^2}(x-x_0)\right]
        }
    \right),
\end{equation}
where $\omega$ is the breather frequency and $x_0$ is the
coordinate of its center.

The case of $K(x) = 1-\varepsilon \delta(x)$ where $\delta(x)$ is
the Dirac delta function and $0 < \varepsilon <1$ is a constant
has been studied in  \cite{Kivshar2008eng}. It has been
demonstrated that in frame of the underformable kink approximation
the impurity acts as a potential and for the chosen sign of
$\varepsilon$ the potential is attractive and hence the soliton
can be localized on the impurity. In the case of deformable kink,
in addition to the oscillatory motion of the kink in the potential
generated by the impurity, a strong modification of the kink
shape, having a resonant character, can take place. The
possibility of excitation of the impurity mode as a result of kink
scattering that results in a considerable change of kink dynamics
has also been considered. For the case of finite size impurity (in
the simplest case $K(x)$ has the form \eqref{Impurity} the
interaction of the kink with the impurity has been analyzed for
both non-deformable and deformable kinks
\cite{PMM2006,Paul1979,Piette2007kinks}.

Let us investigate the relation between the spectra of the
impurity mode and the small-amplitude excitations of
\eqref{Lagrangian}. We take into account that for the
one-dimensional case of \eqref{EqMotion} with $K(x) = 1$ there
exists the vacuum solution $\theta_\pm(x,t)=0$. We look for the
spectrum of the small-amplitude vibrations in the vicinity of this
solution
\begin{equation}
    \label{six}
    \theta(x,t) =\theta_\pm+\delta\theta(x,t), \quad \delta\theta(x,t)\ll
    1.
\end{equation}

Substituting  \eqref{six}  into \eqref{EqMotion} after linearization with respect to $\delta\theta$ ,
one gets the equation
\begin{equation}
    \label{seven}
    L\delta\theta(x) =\omega^2_n\delta\theta(x),
\end{equation}
which is the Schr\"{o}dinger  equation with the operator  $L=-\frac{d^2}{dx^2} +K(x)$, where  $\omega_n$
is the impurity mode frequency. Let us look for the localized
solutions of the Schr\"{o}dinger  equation \eqref{seven}. It is convenient to
introduce the following notations
\begin{equation}
    \label{eight}
    \chi^2=1-\omega^2, \quad   k^2=\omega^2-(1-\Delta K).
\end{equation}

For $K(x)$ defined by \eqref{Impurity} even and odd solutions to \eqref{seven} are possible  \cite{Landau2004eng}
\begin{eqnarray}
    \label{ninea}
    \Psi_+  = \left\{
        \begin{array}{lll}
            A_1^+e^{\chi x}, & x<-W/2 \\
            B_2^{+} \cos kx, & -W/2\le x \le W/2 \\
            B^+_3e^{\chi x}, & x>W/2
        \end{array}
    \right.
\end{eqnarray}
\begin{eqnarray}
    \label{nineb}
    \Psi_-  = \left\{
        \begin{array}{lll}
            A_1^-e^{\chi x}, & x<-W/2 \\
            B_2^{-} \sin kx, & -W/2\le x \le W/2 \\
            B^-_3e^{\chi x}, & x>W/2
        \end{array}
    \right.
\end{eqnarray}

We subject the solution to the condition of smoothness and
continuity at the point $x=W/2$ and obtain the following
dispersion relations for even \eqref{ninea} and  odd \eqref{nineb}
solutions, respectively,
\begin{equation}
    \label{tena}
    \tan\left(k\frac{W}{2}\right)=\frac{\chi}{k},
\end{equation}
\begin{equation}
    \label{tenb}
    \tan\left(k\frac{W}{2}\right)=-\frac{\chi}{k}.
\end{equation}

Since, in the considered case,  $\Delta K$ and $W$ are constant,
equations (\ref{tena},\ref{tenb}) give the possibility to find all
frequencies $\omega$ of the impurity modes in the potential well
of given size. The states with even and odd wave functions
alternate and the first odd solution, that corresponds to the
second localized state, appears when the relation $\pi^2=W^2\Delta
K$ is satisfied.

Let us study numerically the large-amplitude localized impurity
modes excited due to the interaction with a kink for the case
$\Delta K\le 0$. The most interesting case is when the size of the
soliton is of the same order with the size of the impurity because
in this case the soliton shape is strongly affected by the
impurity. To solve the equations of motion \eqref{EqMotion}
numerically we use the iteration method for the explicit scheme.
The following algorithm was applied. Initially we have a SGE kink
(4) moving with a constant speed. Boundary conditions have the
form $\theta(\pm\infty)=0, \pi$; $\theta'(\pm\infty)=0$. We
introduce the mesh for the spatial coordinate and iterate with
respect to time, taking into account the convergence condition for
the explicit scheme, to find the kink position at the next time
step. The characteristics of the nonlinear wave were found from
the numerically constructed function $\theta(x,t)$.

\begin{figure}[t]
    \begin{center}
        \includegraphics[width=0.5\linewidth]{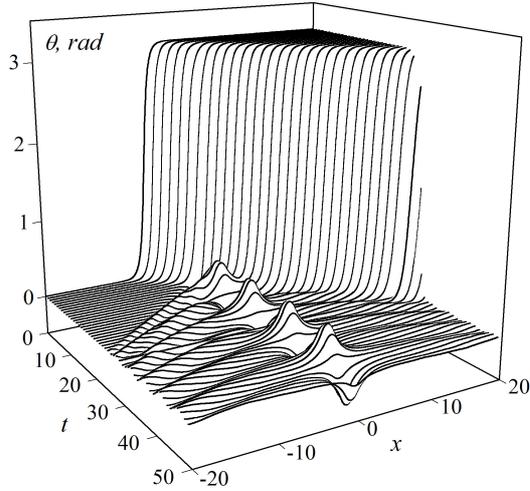}
        %\smallskip \medskip \bigskip \vspace[*]{length}
        \caption{Excitation of the breathing nonlinear excitation
        on the impurity by the moving kink for the case of
        $\Delta K = 1.5$, $W = 1.5$, $v_0 = 0.86$. The impurity
        center is located at $x=0$.}
        \label{fig:1}
    \end{center}
\end{figure}

\begin{figure}[t]
    \begin{center}
        \includegraphics[width=0.33\linewidth]{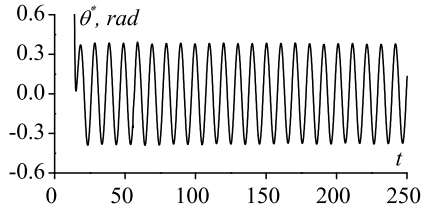}
        \caption{Time dependence of the field value at the center of the impurity
        showing the oscillation of the breather excited by the passing kink as
        shown in Fig.\ref{fig:1}. The numerically obtained result overlaps with
        that calculated from (\ref{eleven}) for $x=0$ and
        $A=0.4,\, \omega=0.616,\, \alpha=4.5\times 10^{-5},\, \gamma=1,\, t_0=4.5$.}
        \label{fig:2}
    \end{center}
\end{figure}

The numerical experiments have demonstrated that the kink passing
the impurity excites the bell-shaped nonlinear oscillatory wave,
see Fig. \ref{fig:1}. We found that this oscillatory mode can be
well fitted by the expression
\begin{eqnarray}
    \label{eleven}
    \theta^{*}(x,t) = A \exp( -\alpha(t-t_0) )
                \arctan \left(
                    \frac{\sqrt{1-\omega^2}\sin[\omega\gamma(t-t_0)]}
                    {\omega\,\sech \left(\gamma \sqrt{1-\omega^2}\,x\right)}
                \right),
\end{eqnarray}
which is a stationary breather (\ref{Breather}) in the presence of
damping with the coefficient $\alpha$ and with an additional
fitting parameter $\gamma$. In Fig. \ref{fig:2} we compare the
numerically found time evolution of the field value at the center
of the impurity $(x=0)$ with that found from the approximating
formula (\ref{eleven}) with the fitting parameters $A=0.4,\,
\omega=0.616,\, \alpha=4.5\times 10^{-5},\, \gamma=1,\, t_0=4.5$.
The numerical result and the result obtained from the fitting
formula (\ref{eleven}) overlap. Damping of the breather is due to
slow radiation of small-amplitude extended waves that can be seen
in Fig. \ref{fig:1}.

\begin{figure}[t]
    \begin{minipage}[t]{0.47\linewidth}
        \center{\includegraphics[width=0.66\linewidth]{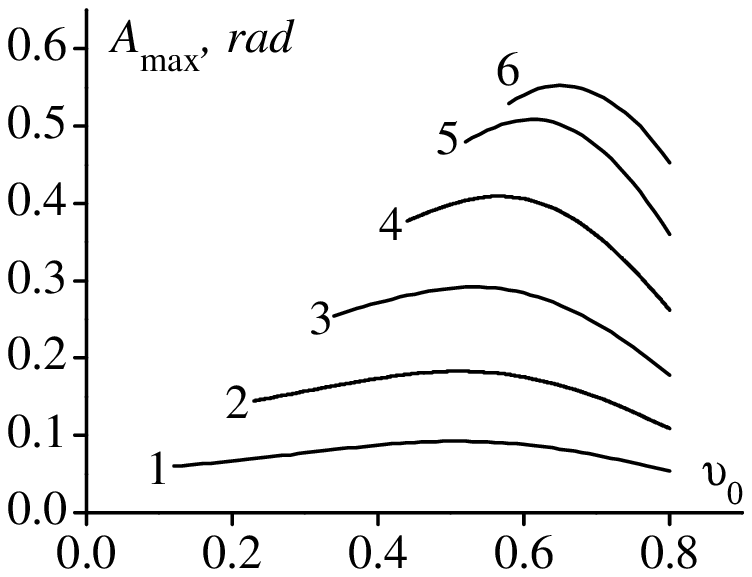} \\ (a)}
    \end{minipage}
    \hfill
    \begin{minipage}[t]{0.47\linewidth}
        \center{\includegraphics[width=0.66\linewidth]{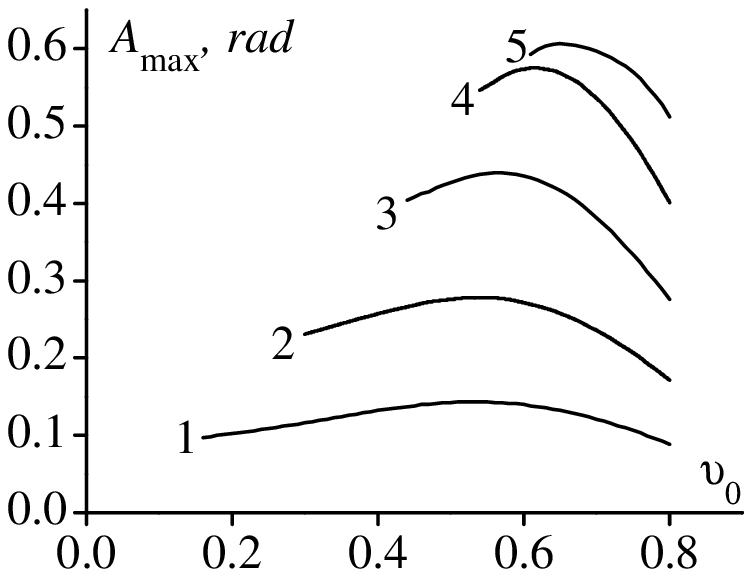} \\ (b)}
    \end{minipage}
    \caption{Breather amplitude $A_{\max}$ measured at the impurity center
    as the function of kink velocity $v_0$ for (a) $W = 1$ and (b) $W = 1.5$.
    Curves 1 to 6 are for $\Delta K= \{ 0.5,\, 0.75,\, 1,\, 1.25,\, 1.5,\, 1.75\}$, respectively.}
\label{fig:3}
\end{figure}

Amplitude of the breather excited by the passing kink depends on
the kink velocity $v_0$ (see Fig. \ref{fig:3}) and the curve has a
maximum with the value dependent on the impurity parameters
$\Delta K$ and $W$.

The amplitude of the excited breather also depends on $\Delta K$
and $W$ and it vanishes when $\Delta K \to 0$, $W \to 0$, i.e., in
the absence of impurity.

\begin{figure}[t]
    \begin{minipage}[t]{0.47\linewidth}
        \center{\includegraphics[width=0.66\linewidth]{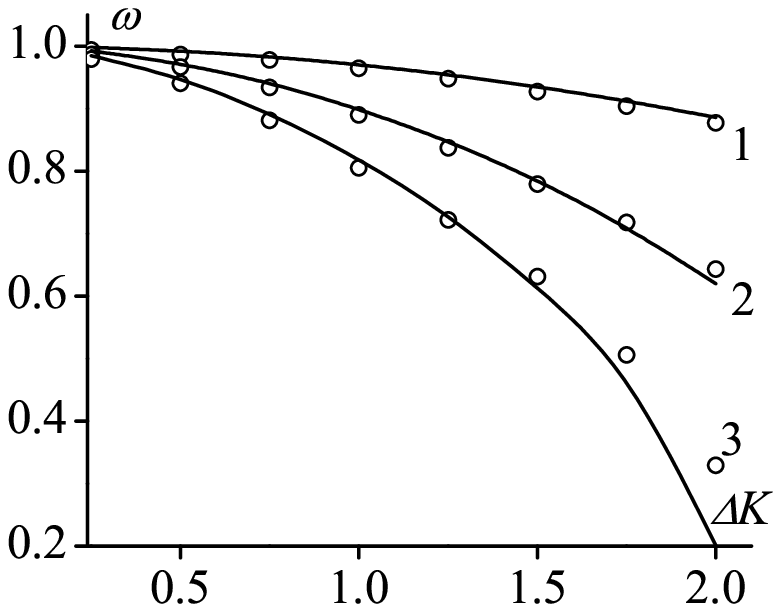} \\ (a)}
    \end{minipage}
    \hfill
    \begin{minipage}[t]{0.47\linewidth}
        \center{\includegraphics[width=0.66\linewidth]{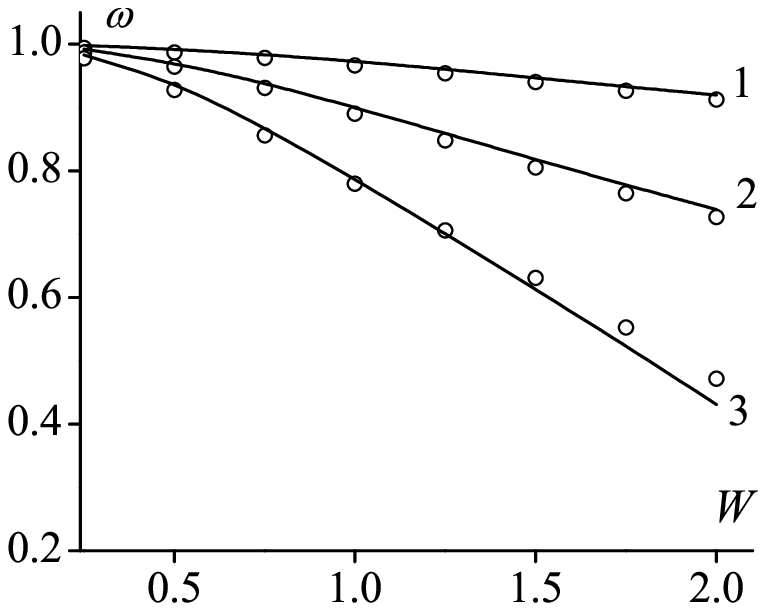} \\ (b)}
    \end{minipage}
    \caption{Dependence of the breather oscillation frequency on the
    parameters $\Delta K$ (a) and $W$ (b). The solid line corresponds
    to the frequency calculated from \eqref{tena}, while scattered data
    was obtained by numerical integration of \eqref{1Dcase}. In (a)
    curves 1 to 3 correspond to $W=0.5$, $W=1$, and $W=1.5$,
    respectively, while in (b) to $\Delta K=0.5$, $\Delta K=1$, and
    $\Delta K=1.5$, respectively.}
\label{fig:4}
\end{figure}

Breather frequency $\omega_B$ is practically independent of the
kink velocity $v_0$, while it depends on $\Delta K$ and $W$ as
shown in Fig. \ref{fig:4}.

\begin{figure}[t]
    \begin{minipage}[t]{0.47\linewidth}
        \center{\includegraphics[width=0.66\linewidth]{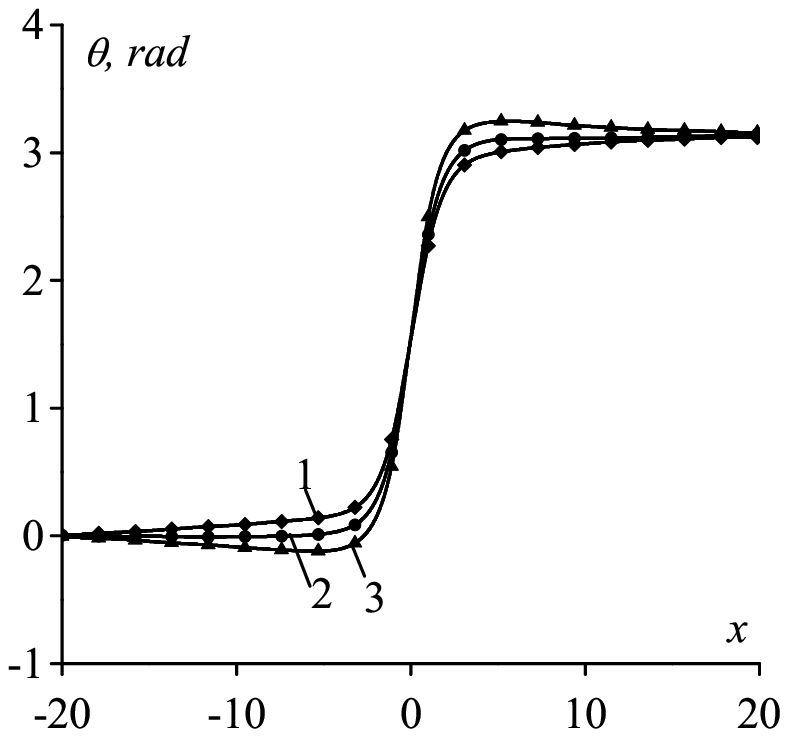} \\ (a)}
    \end{minipage}
    \hfill
    \begin{minipage}[t]{0.47\linewidth}
        \center{\includegraphics[width=0.66\linewidth]{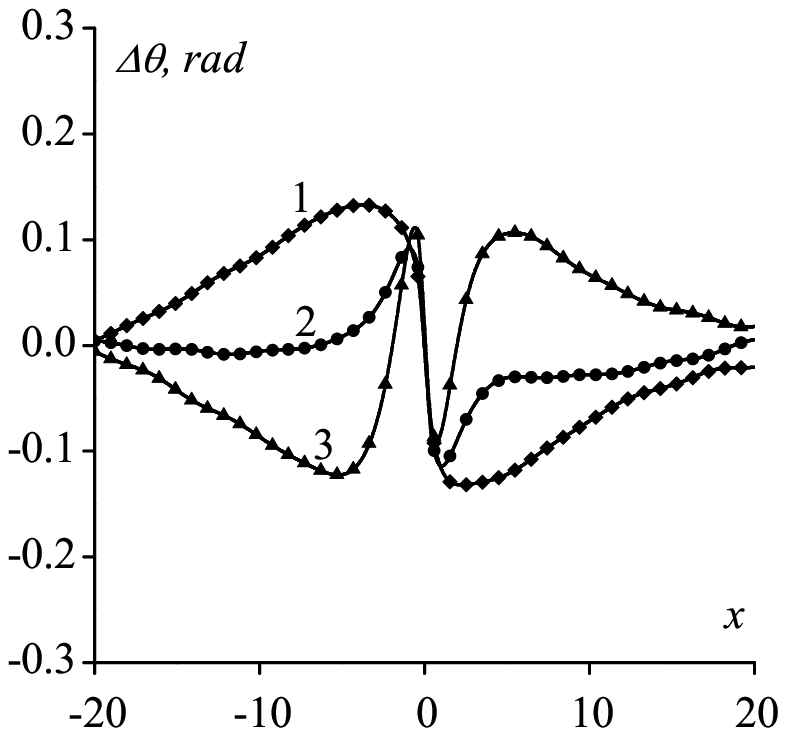} \\ (b)}
    \end{minipage}
    \caption{(a) Wobble kink profile and (b) its difference from the
    kink solution at different times, 1 -- $t=1179.37$, 2 -- $t=1180.87$, and 3 --$t=1182.37$,
    for the case $W=1$, $\Delta K=1.2$ and initial kink velocity $v_0=0.2$.
    The impurity center is at $x^{*}=-0.5$.}
\label{fig:5}
\end{figure}

It can be seen that for  $\Delta K \to 0$, $W \to 0$ the breather
frequency tends to unity but the frequencies of kink's
translational and pulsation (oscillation of kink wigth) modes
excited due to the interaction with impurity tend to zero
\cite{PMM2006}. This behavior can be easily understood taking into
account that breather energy scales as $E\sim(1-\omega^2_B)^{1/2}$
\cite{Scott2004} meaning that for vanishing size of the impurity
the energy and the amplitude of the breather also vanish. It can
also be seen that the linear approximation for the first even
solution described above gives a good description of the excited
breathers in the range of small amplitudes because the frequencies
calculated from the analytical expression \eqref{tena} (solid lines
in Fig. \ref{fig:4}) coincide with the frequencies found
numerically by solving \eqref{EqMotion}. Excitation of the first
odd mode (the second main state) will be discussed for the case of
kink pinning by the impurity. In this case, as it was already
mentionaed, pulsation and translational modes can be excited on
the kink  \cite{PMM2006}. In the case when kink pulsation mode
frequency $\omega_{pulse}<1$, after a transient period a state is
formed that can be described with a good accuracy by the kink-type
solution. Therefore, in our case the most interesting is the range
of parameters $\Delta K$ and $W$ where $\omega_{pulse}\to 1$ and
the excited nonlinear wave (see Fig. \ref{fig:5}a) differs
considerably from the kink solution (see Fig. \ref{fig:5}b).
Algebraic difference of the solutions has the form dramatically
different from that shown in Fig. \ref{fig:1} and it is analogous
to the form of the odd solutions to the Schr\"{o}dinger equation.

\begin{figure}[t]
    \begin{center}
        \includegraphics[width=0.33\linewidth]{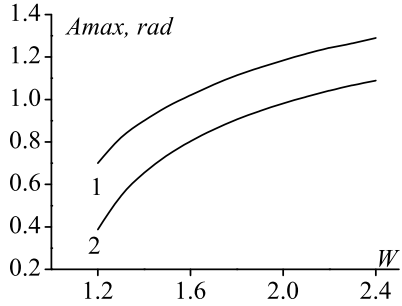}
        %\smallskip \medskip \bigskip \vspace[*]{length}
        \caption{The dependence of the stabilized soliton amplitude
        on the impurity width $W$ at fixed $\Delta K= 2.8$.
        Curve 1 gives numerical result, while curve 2 corresponds
        to \eqref{GrindEQ__12_}.}
        \label{fig:6}
    \end{center}
\end{figure}

\begin{figure}[t]
    \begin{center}
        \includegraphics[width=0.33\linewidth]{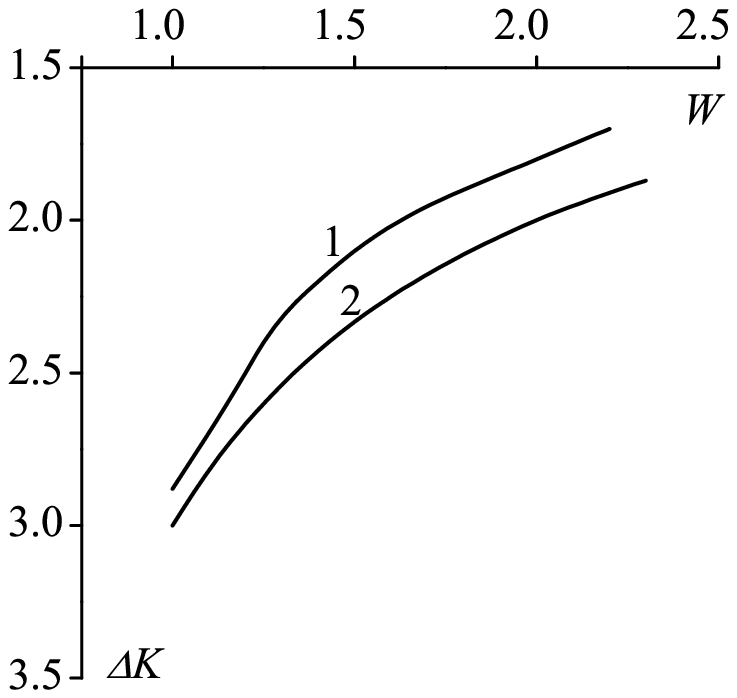}
        %\smallskip \medskip \bigskip \vspace[*]{length}
        \caption{Space of impurity parameters showing the range where passing
        kink excites a breather (above the lines 1 and 2), or a soliton (below
        the lines 1 and 2). Line 1 was found numerically and line 2 is given by
        \eqref{twelve}.}
        \label{fig:7}
    \end{center}
\end{figure}

One should also take into account that for the case of
large-amplitude nonlinear waves obtained numerically, a three-kink
solution can be not just a linear sum of the kink and breather
solutions but a modified solution that takes into account kink
vibrations described by the wobble SGE solution
\cite{Ferreira2008,Kalberman2004}. For yet increasing values of
$\Delta K$ and $W$ let us take into account that, as it has been
shown in \cite{PMM2008eng}, at the critical value of parameters,
\begin{equation}
    \label{twelve}
    KW=2,
\end{equation}
in the vicinity of the impurity a stable static soliton can exist
whose amplitude can be estimated as
\begin{equation}
    \label{GrindEQ__12_}
    \cos A_{\max } =2/KW,
\end{equation}
where $K=1-\Delta K$. Numerical results show that for sufficiently
large $\Delta K$ and $W$ after the passing of kink, in the
vicinity of the impurity a soliton is formed. The dependence of
soliton amplitude on $\Delta K$ and $W$ (see Fig. \ref{fig:6}) can
be approximately expressed as $\cos A=1.8/KW$. In Fig. \ref{fig:7}
the ranges of the impurity parameters where the breather and the
soliton can exist are presented and for comparison the curve
defined  by \eqref{twelve} is plotted.

\section{Two-dimensional case}
\label{2Dcase}

For definiteness let us take the impurity in the following form
\cite{Bratsos2007}
\begin{eqnarray}
\label{GrindEQ__13_}
    K(x,y)= \left\{
        \begin{array}{ll}
            1, & x<x_{1} ,x>x_{2}, y<y_{1} ,y>y_{2} \\
            1-\Delta K, & x_{1} \le x\le x_{2}, y_{1} \le y\le y_{2}
         \end{array}
    \right.
\end{eqnarray}
where $W_{x} =x_{1}-x_{2}$ and $W_{y} =y_{1}-y_{2}$ are the
parameters that specify the size of the impurity. Similarly to the
one-dimensional case, Eq. \eqref{EqMotion} was integrated with the
use of an explicit scheme. A uniform mesh with the step $\xi $ was
introduced for $x$ and $y$ coordinates,
\begin{equation}
\label{GrindEQ__13aa_}
    x_{i} =\xi i,\,\, i=-N_{x} ,...,N_{x}, \quad  y_{j} =\xi j,\,\,  j=-N_{y}
    ,...,N_{y},
\end{equation}
and for the time $t$ the mesh with the time step $\tau$ was employed,
\begin{equation}
\label{GrindEQ__13bb_}
    t_{n} =\tau n, \,\, n=0,1,...,N_{t},
\end{equation}
where  $N_{x}$, $N_{y}$, $N_{t}$ are integers. For $N_{x}$ and
$N_{y}$ we took the values from 512 to 2048. It was confirmed that
the number of mesh points for the spatial coordinates does not
affect the main numerical results. Initially we have a SGE kink
\eqref{Kink} moving with constant velocity. The boundary conditions
for the $x$-coordinate have the form $\theta (\pm N_{x}
\xi,y)=\theta _{0} (\pm N_{x} \xi)$; $\theta '(\pm N_{x}
\xi,y)=\theta '_{0} (\pm N_{x} \xi)$ and for the $y$-coordinate
free edges are simulated.

\begin{figure}[t]
    \includegraphics[width=1\linewidth]{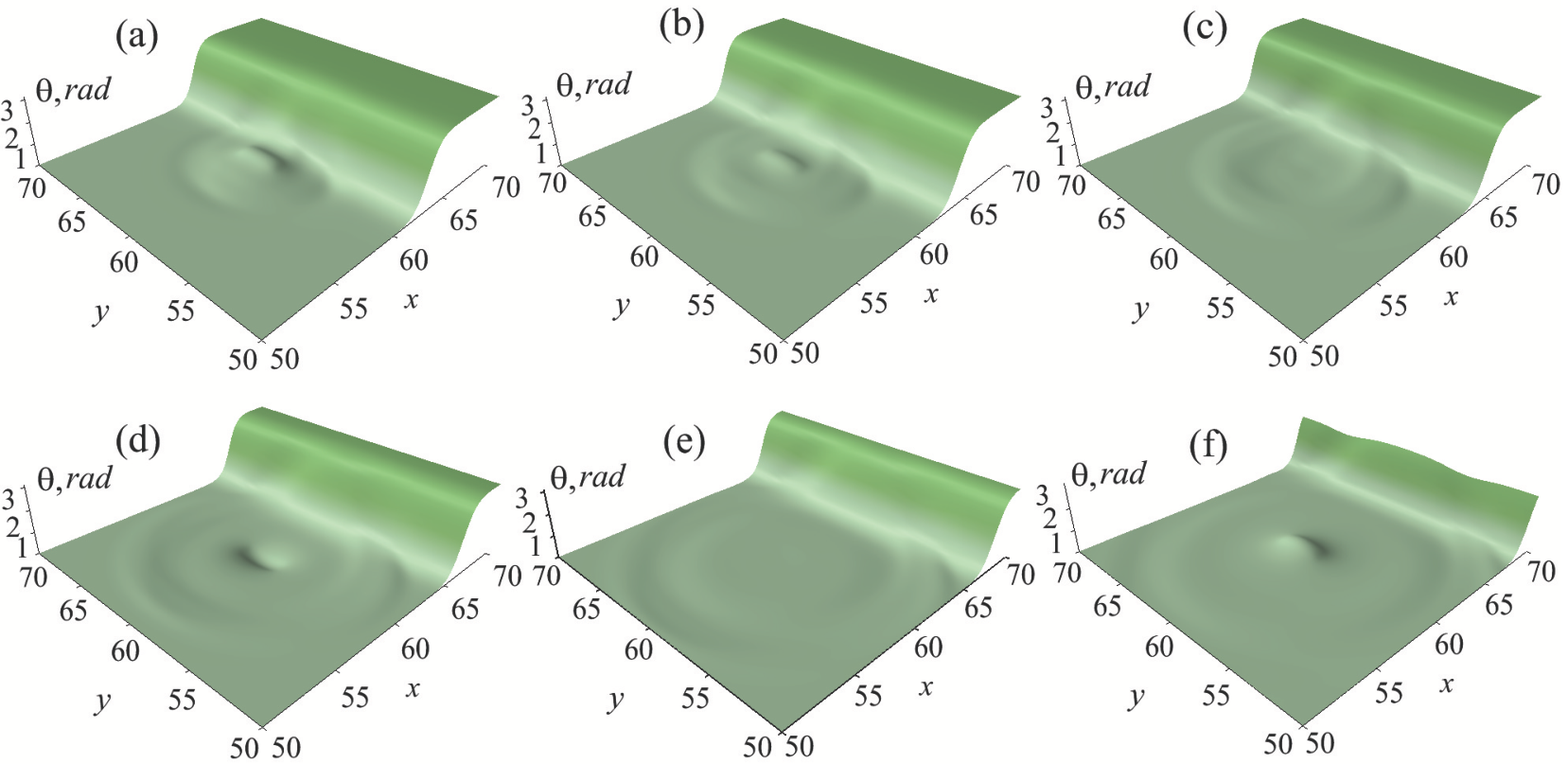}
    \caption{The excitation and evolution of a breathing pulson
    for the case $W_x=1$, $W_y=3$, and $\Delta K=2$. (a)~$t=11.55$, (b)~$t=12.6$, (c)~$t=13.65$,
    (d)~$t=15.75$, (e)~$t=17.43$, anf (f)~$t=19.32$.
    The impurity center coordinates are $x^{*}=y^{*}=60$.}
    \label{fig:8}
\end{figure}

\begin{figure}[p]
    \begin{center}
            \includegraphics[width=0.5\linewidth]{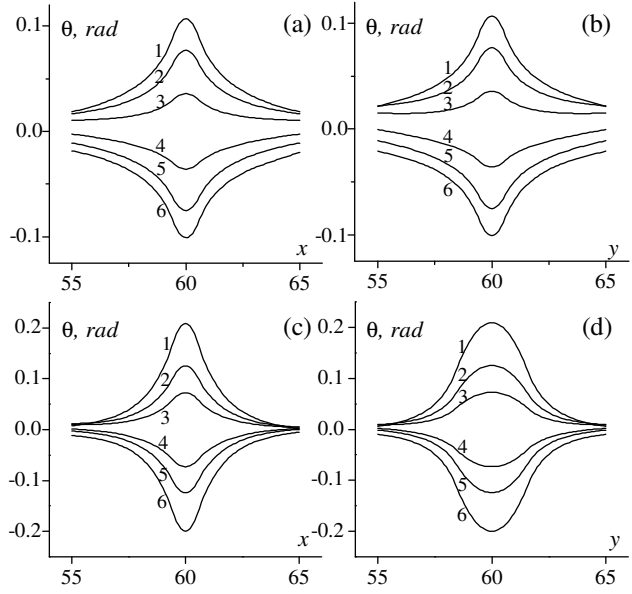}
            %\smallskip \medskip \bigskip \vspace[*]{length}
            \caption{Snapshots of the functions $\theta(x,y^{*},t)$ (a,c)
            and $\theta(x^{*},y,t)$ (b,d) for the case
            $\Delta K=2$, $v_0=0.57$. In (a,b) the impurity is symmetric,
            $W_x=W_y=1$, and the curves 1 to 6 are for
            $t=\{28.2,\,29.04,\,29.5,\,30.24,\,30.72,\,31.5\}$. In (c,d) the impurity is asymmetric,
            $W_x=1$, $W_y=3$, and the curves 1 to 6 are for
            $t=\{44.28,\,45.6,\,45.96,\,46.5,\,46.86,\,48.18\}$. The impurity center
            coordinates are $x^{*}=y^{*}=60$.}
            \label{fig:9}
    \end{center}
\end{figure}

\begin{figure}[p]
    \begin{center}
            \includegraphics[width=0.45\linewidth]{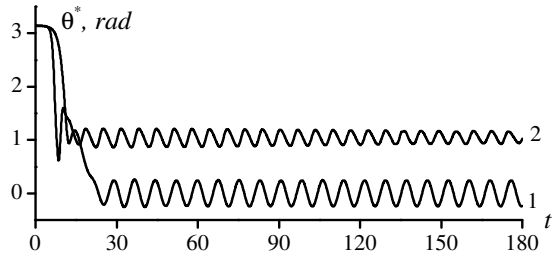}
            %\smallskip \medskip \bigskip \vspace[*]{length}
            \caption{The function $\theta(x^{*},y^{*},t)$ for the case
            $v_0=0.85$, $W_x=1$, $W_y=3$, $\Delta K=2$ (curve 1)
            and $\Delta K=5$ (curve 2).}
            \label{fig:10}
    \end{center}
\end{figure}

\begin{figure}[p]
    \begin{minipage}[t]{0.47\linewidth}
        \center{\includegraphics[width=0.66\linewidth]{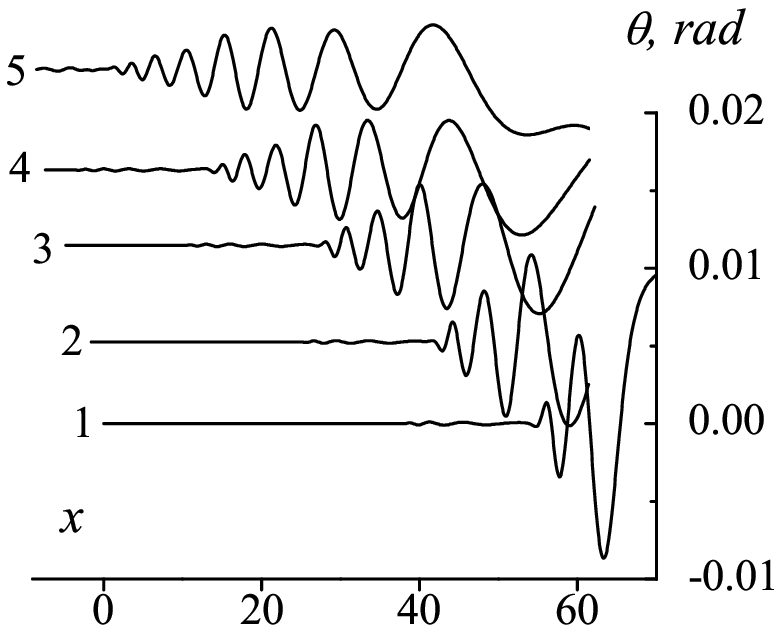} \\ (a)}
    \end{minipage}
    \hfill
    \begin{minipage}[t]{0.47\linewidth}
        \center{\includegraphics[width=0.66\linewidth]{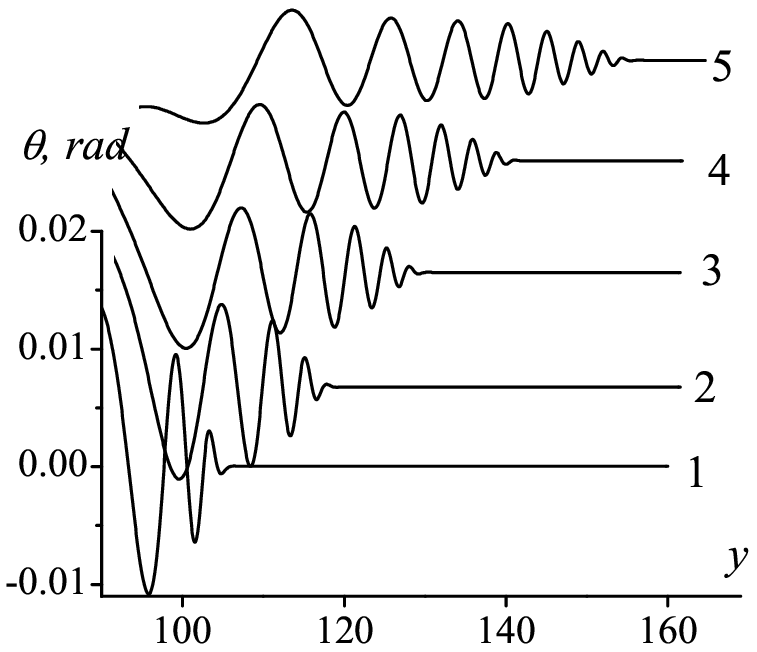} \\ (b)}
    \end{minipage}
    \caption{Snapshots of $\theta(x,y^{*},t)$ (a) and $\theta(x^{*},y,t)$ (b)
    for $W_x=1$, $W_y=1$, $\Delta K=2$. Curves 1 to 5 correspond
    to $t=\{42,\,54,\, 66,\, 78, \,90\}$.}
	\label{fig:11}
\end{figure}

Let us consider the case of $\Delta K> 0$ when, as it was shown
above, nonlinear waves localized at the impurity are excited due
to the interaction with the kink. In Fig. \ref{fig:8} the process
of kink-impurity interaction is presented and one can see that the
nonlinear localized wave radiating extended waves is excited. The
wave is called here breathing pulson. If $W_{x} =W_{y}$ then the
breathing pulson is symmetric with respect to $x$ and $y$ axis
(Fig. \ref{fig:9}a,b) and the symmetry is lost for $W_{x} \neq
W_{y}$ (Fig. \ref{fig:9}c,d). Firstly, the excited localized wave
has a bell shape and later ($t>20$, curve 1 in Fig. \ref{fig:10})
the periodic oscillations can be seen at the center of the
impurity having coordinates $\left(x^{*} ,y^{*} \right)$. With
increase in the impurity size the oscillation frequency reduces.
The amplitude decreases with time owing to the radiation of
extended waves. It should be noted that the extended waves cannot
be described by the harmonic function because they have a
nonlinear nature (see Fig. \ref{fig:11}). The breathing pulson can
be regarded as a long-lived nonlinear excitation.

It is well-known that SGE supports the 2D solution called pulson
describing long-lived spatially localized vibrations
\cite{Dodd1982,Qing2010,Koutvitsky2005}. The numerically found
breathing pulson can be well approximated by the expression
\begin{eqnarray}
    \label{GrindEQ__14_}
    \theta (r,t)= A \arctan \left[
        \frac{\sqrt{1-\omega_{P} ^{2} } }{\omega_{P} }
        \sech\left(\sqrt{1-\omega_{P} ^{2}} Br\right)
        \sin [\omega_{P} (t-t_{0} ) ],
    \right]
\end{eqnarray}
where $r=\sqrt{({x / \Delta _{x}})^{2} +({y / \Delta _{y}
})^{2}}$, with $\Delta _{x}$ and $\Delta _{y}$ being the pulson
width along $x$ and $y$ axis, respectively. For example, for the
curve 1 in Fig. \ref{fig:10} one should set in
(\ref{GrindEQ__14_}) $r=0$ and $A=0.42$, $\omega_{P}=0.85$,
$B=4.8$, $t_{0} =2.5$, $\Delta_{x}=2.4$, and $\Delta_{y}=3.4$.

\begin{figure}[p]
    \begin{minipage}[t]{0.47\linewidth}
        \center{\includegraphics[width=0.66\linewidth]{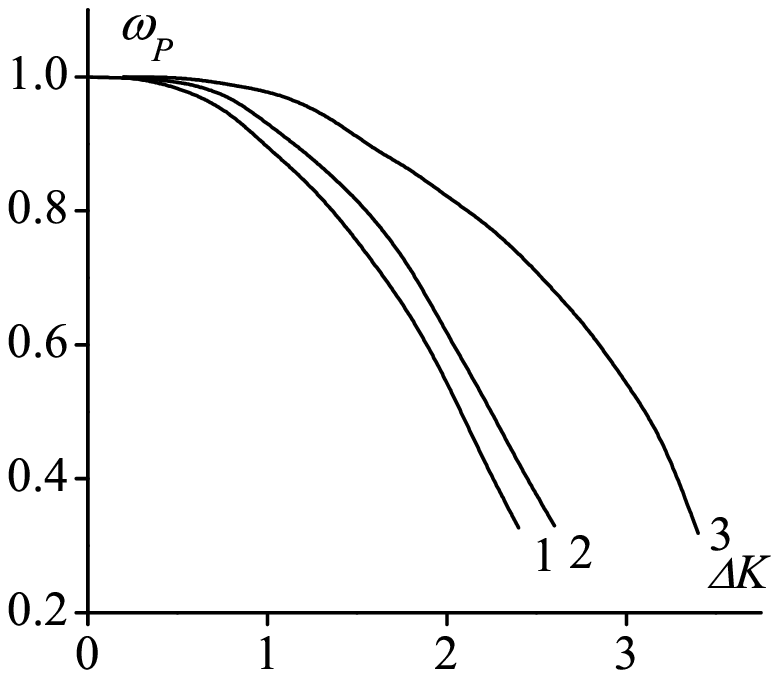} \\ (a)}
    \end{minipage}
    \hfill
    \begin{minipage}[t]{0.47\linewidth}
        \center{\includegraphics[width=0.66\linewidth]{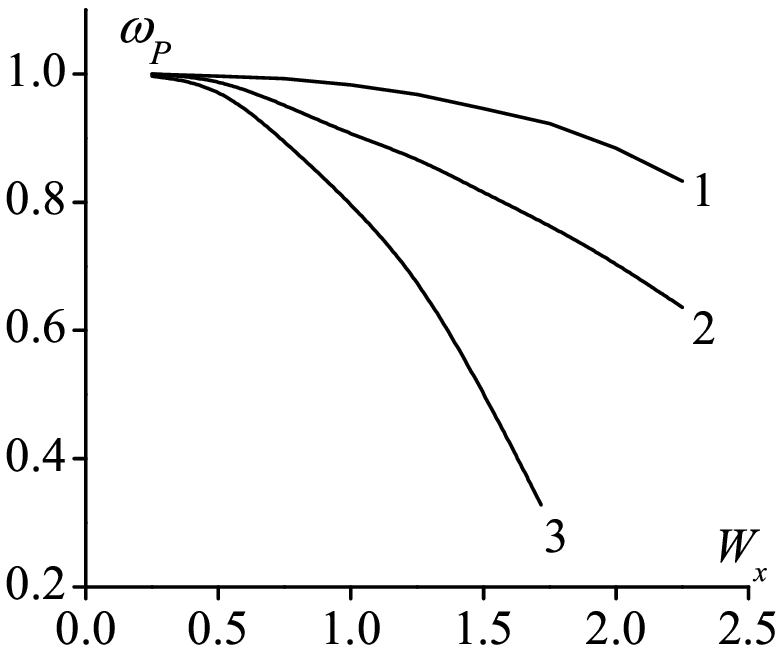} \\ (b)}
    \end{minipage}
    \caption{The dependence of the breathing pulson oscillation frequency $\omega_p$
    (a) on the parameter $\Delta K$ for the case $W_y=3$ and $W_x=1$ (curve 1),
    $W_x=2$ (curve 2), and $W_x=3$ (curve 3); (b) on the parameter $W_x$
    for the case $W_y=1$ and $\Delta K=2$ (curve 1), $\Delta K=3$ (curve 2), and $\Delta K=4$ (curve 3).
    Kink initial velocity is $v_0=0.57$.}
	\label{fig:12}
\end{figure}

\begin{figure}[p]
    \begin{center}
        \includegraphics[width=0.33\linewidth]{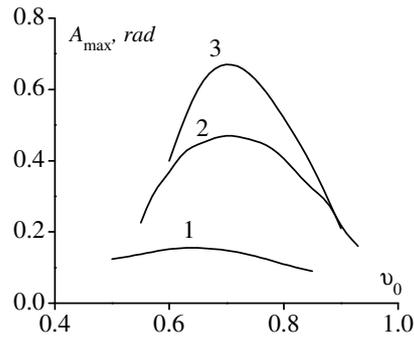}
        \caption{The dependence of the breathing pulson maximum amplitude
        $A_{\max}$ at the impurity center on the kink velocity $v_0$
        for the case $W_x=1$, $\Delta K=2$, and $W_y=1$ (curve 1),
        $W_y=2$ (curve 2), $W_y=3$ (curve 3).}
        \label{fig:13}
    \end{center}
\end{figure}

Our numerical results show that the breathing pulson frequency
$\omega_{P}$ [determined from $\theta (x^{*}, y^{*}, t)$] does not
depend on kink velocity $\upsilon _{0}$, but it is a function of
$W_{x}$, $W_{y}$ and $\Delta K$. In Fig. \ref{fig:12} the
dependence of $\omega_{P}$ on the parameters $\Delta K$ and
$W_{y}$ is shown. It can be seen that with decrease in the
impurity size the breathing pulson frequency (similar to what was
observed for the breather) tends to unity. Dependence of
$\omega_{P}$ on $K=1-\Delta K$ can be approximately given by
$\omega_{P} ={(a(|K^{*}|+K))^{q} / \sqrt{1+(a(|K^{*}|+K))^{2q} }
}$, where $a$ is a constant, $q\approx 2$, $K^{*}$ is the smallest
value of $K$ when breathing pulson is still formed in the vicinity
of the impurity. Dependence of $\omega_{P}$ on $W_{y}$ (as well as
on $W_{x}$) can be approximately expressed as
$\omega_{P}=1-(bW_{y} )^{p} / \sqrt{1+(bW_{y} )^{2p} }$, where $b$
is a constant and $p\approx 2$. Maximal breathing pulson
amplitude, $A_{\max}$, as the function of the kink velocity
$\upsilon _{0}$ is presented in Fig. \ref{fig:13} and, similarly
to the one-dimensional case, it has a maximum. With decrease in
the impurity size $A_{\max}$ vanishes. The dependence
$A_{\max}(\Delta K)$ for small values of $\Delta K$ is close to
linear. The dependence $A_{\max}(W_{y})$ for $W_{y}<1$ is close to
linear and for large $W_{y}$ it saturates. Note that for fixed
$\upsilon _{0}$ maximal value of $A_{\max}$ strongly depends on
$\Delta K$, $W_{x}$ and $W_{y}$.

\begin{figure}[p]
    \includegraphics[width=1\linewidth]{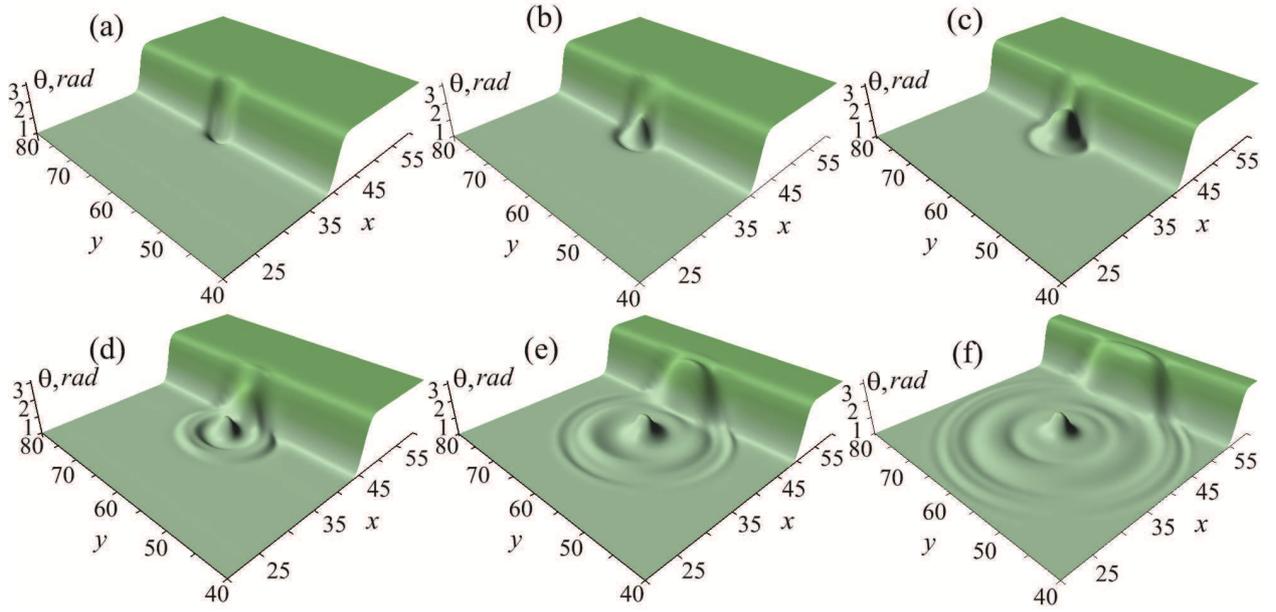}
    \caption{Excitation and evolution of the breathing 2D soliton
    for the case $W_x=1$, $W_y=3$, $\Delta K=5$. Panels (a) to (f)
    correspond to $t=\{8,\,9.4,\,11,\,14,\,20,\,26\}$.
    The impurity center coordinates are $x^{*}=y^{*}=60$.}
    \label{fig:14}
\end{figure}

\begin{figure}[p]
    \begin{center}
        \includegraphics[width=0.5\linewidth]{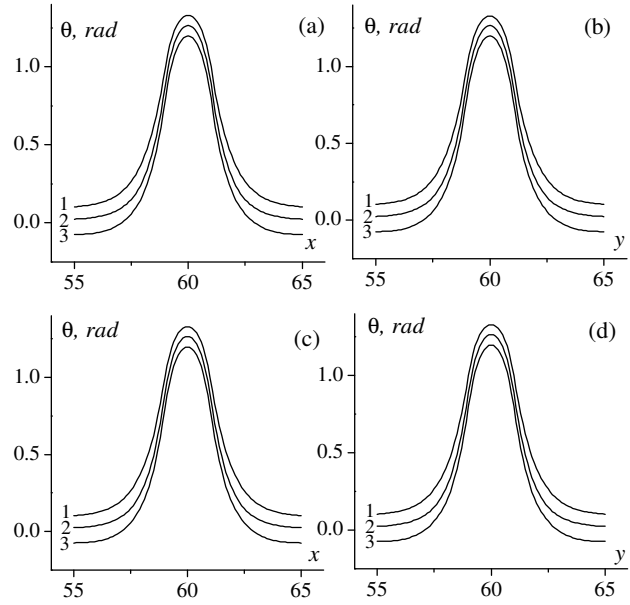}
        \caption{Snapshots of the functions $\theta(x,y^{*},t)$ (a,c)
        and $\theta(x^{*},y,t)$ (b,d) for the case (a,b) $W_x=2$, $W_y=2$, and
        $t=30.18$ (curve 1), $t=31.8$ (curve 2), $t=33.72$ (curve3);
        (c,d) $W_x=2$, $W_y=3$, and $t=30.78$ (curve 1), $t=32.4$ (curve 2), $t=34.02$ (curve 3).
        $\Delta K= 5$, $v_0=0.85$. The impurity center coordinates are $x^{*}=y^{*}=60$.}
        \label{fig:15}
    \end{center}
\end{figure}

For increasing values of the parameters $\Delta K$, $W_{x}$ and
$W_{y}$, after the kink passes the impurity, a localized nonlinear
wave called here breathing 2D soliton is excited on the impurity
(see Fig. \ref{fig:14}). The breathing 2D soliton shape depends on
the parameters $W_{x}$ and $W_{y}$ and it can be symmetric (Fig.
\ref{fig:15}a,b) or asymmetric (Fig. \ref{fig:15}c,d). Our
numerical results suggest that the breathing 2D soliton is a
long-lived excitation with amplitude slowly decreasing in time.
The breathing 2D soliton cannot be described by the direct sum of
2D soliton solution and pulson solution to SGE. On the other hand,
the breathing 2D soliton can be well approximated by the
expression
\begin{eqnarray}
    \label{GrindEQ__15_}
    \theta (r,t)= \arctan \left( \frac{
        A_{0} +A_{1} \sin [\omega_{S} (t-t_{0} )]
    }{
        \omega_{S}^{-1}\sqrt{1-\omega_{S} ^{2} }
        \cosh(\sqrt{1-\omega_{S} ^{2} }Br)
    } \right),
\end{eqnarray}
where $r=\sqrt{({x / \Delta _{x} })^{2} +({y / \Delta _{y} })^{2}
}$, $B=B_{0}-B_{1} \sin [\omega_{S} (t-t_{0} )]$, $\omega_{S}$ is
breathing 2D soliton frequency, $A_{0}$, $A_{1}$, $B_{0}$, $B_{1}$
are the breathing 2D soliton parameters that depend on the
impurity parameters, $\Delta_{x}$, $\Delta_{y}$ are the
characteristic widths of the soliton along $X$ and $Y$ axis,
respectively. For example, the curve 2 in Fig. \ref{fig:10} is
well fitted by (\ref{GrindEQ__15_}) with $r=0$ and $A_{0}=0.63$,
$A_{1}=0.13$, $B_{0}=9$, $B_{1}=1$, $\omega_{S}=0.94$, $t_{0}=0$,
$\Delta_{x}=1.8$, and $\Delta_{y} =3.2$.

\begin{figure}[p]
    \begin{minipage}[t]{0.47\linewidth}
        \center{\includegraphics[width=0.66\linewidth]{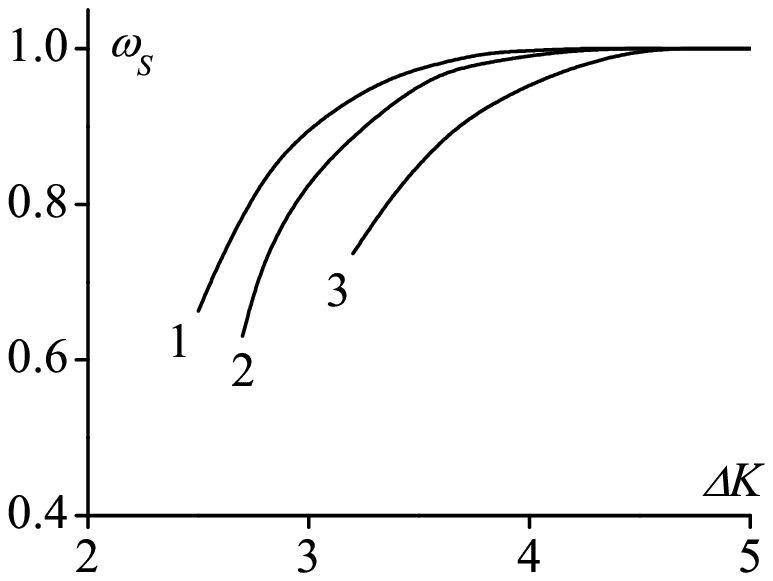} \\ (a)}
    \end{minipage}
    \hfill
    \begin{minipage}[t]{0.47\linewidth}
        \center{\includegraphics[width=0.66\linewidth]{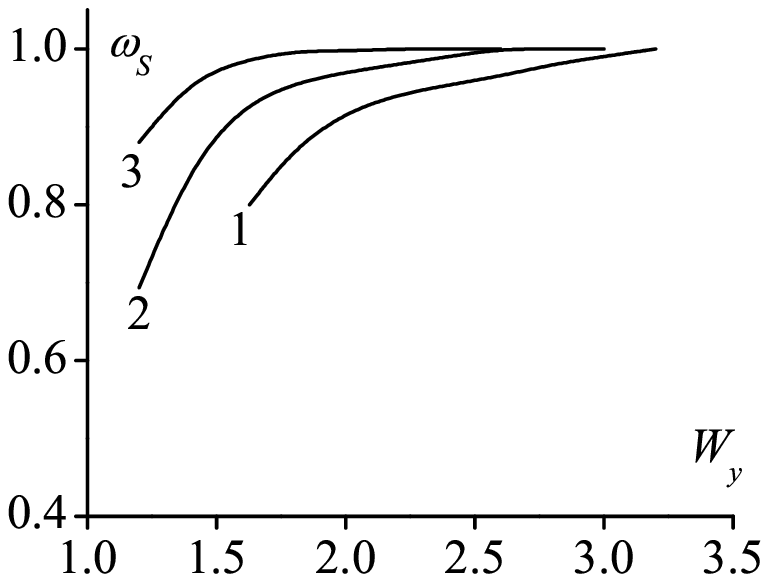} \\ (b)}
    \end{minipage}
    \caption{The dependence of the breathing 2D soliton oscillation frequency $\omega_s$
    (a)~on~the parameter $\Delta K$ for the case $W_y=2$ and $W_x=2$ (curve 1),
    $W_x=3$ (curve 2), $W_x=4$ (curve 3); and (b)~on~the parameter $W_y$
    for the case $W_x=1$ and $\Delta K=5$ (curve 1), $\Delta K=5.5$ (curve 2),
    $\Delta K=6$ (curve 3). Initial kink velocity is $v_0=0.85$.}
\label{fig:16}
\end{figure}

\begin{figure}[p]
    \begin{center}
            \includegraphics[width=0.33\linewidth]{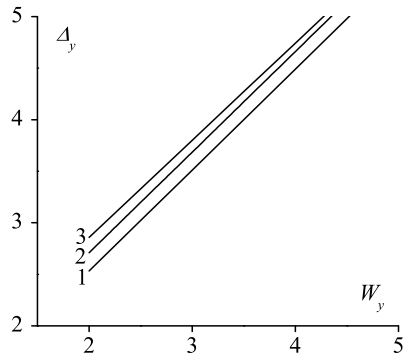}
            %\smallskip \medskip \bigskip \vspace[*]{length}
            \caption{The breathing 2D soliton width $\Delta_{y}$ as the
            function of the parameter $W_y$ for $\Delta K=5$, $v_0=0.85$,
            and $W_x=2$ (curve 1), $W_x=3$ (curve 2), $W_x=4$ (curve 3).}
            \label{fig:17}
    \end{center}
\end{figure}

Our numerical results have demonstrated that the breathing 2D
soliton frequency, $\omega _{S}$, similarly to what was observed
for the pulson, does not depend on the initial kink velocity but
is a function of the parameters  $W_{x}$, $W_{y}$ and $\Delta K$
(see Fig. \ref{fig:16}). It can be seen that $\omega _{S}$ tends
to unity for increasing $W_{x}$, $W_{y}$ and $\Delta K$. The
dependence of $\omega _{S}$ on $K=1-\Delta K$ can be approximated
by the expression
$\omega _{S} ={ (c(|K^*|-K))^{q} / \sqrt{1+(c(|K^*|-K))^{2q} } }$,
where $q\approx 6$, $c$ is a constant, and $K^{*}$ is the largest
value of $K$ at which the breathing 2D soliton can be formed in
the impurity region. The dependence of $\omega _{S}$ on $W_{y}$
(as well as on $W_{x}$) can be approximately given by
$\omega _{S} ={(aW_{y} )^{p} / \sqrt{1+(aW_{y} )^{2p} } }$,
where $p\approx 3$, and $a$ is a constant. In Fig. \ref{fig:17}
for the breathing 2D soliton we present the dependence of $\Delta
_{y}$ on the parameters $W_{y}$. It can be seen that $\Delta _{y}$
depends almost linearly on $W_{y}$. The relation between
parameters describing the size of impurity and the width of the
breathing 2D soliton can be approximately written as
\begin{eqnarray}
    \label{GrindEQ__16_}
    \frac{\Delta _{x}^{2} }{ W_{x}^{2} } +\frac{\Delta _{y}^{2} }{W_{y}^{2} } =
    R^{2},
\end{eqnarray}
where for the symmetric impurity region $R=2^{3/4}$, and, in
general, parameter $R$ depends on the parameters $\Delta K$,
$W_{x}$ and $W_{y}$.

\begin{figure}[p]
    \begin{center}
            \includegraphics[width=0.33\linewidth]{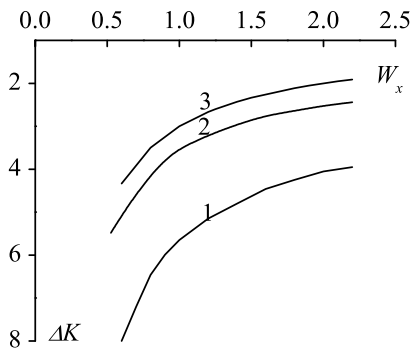}
            %\smallskip \medskip \bigskip \vspace[*]{length}
            \caption{The regions of the impurity parameters allowing
            the existence of the breathing pulson (above the lines 1, 2 and 3),
            and the breathing 2D soliton (below the lines 1, 2 and 3).
            The lines 1 and 2 and 3 present numerical results for $W_y=1$
            and $W_y=3$, respectively, while line 3 is plotted with the help of \eqref{twelve}.}
            \label{fig:18}
    \end{center}
\end{figure}

In Fig. \ref{fig:18} we plot the regions of the parameters $\Delta
K$, $W_{x}$ and $W_{y}$ where different localized nonlinear
excitations exist. It can be seen that the increase in $W_{y}$
shifts the critical curves toward the smaller values of the
parameters $\Delta K$ and $W_{x}$.

\section{Conclusions}
\label{Conclusions}

Using analytical and numerical methods we examined the dynamics of
the sine-Gordon equation kinks passing through the attractive
impurity. The cases of extended one-dimensional and
two-dimensional impurities were studied.

By linearizing the sine-Gordon equation we obtained the dispersion
relations for the small-amplitude localized impurity modes. Using
numerical methods we showed the possibility of excitation by the
passing kink of the first even and odd modes of high-amplitude
localized on the impurity modes. The obtained numerically
dispersion relations in the case of low oscillation amplitudes are
in good agreement with the results of analytical calculations.

For the case of two-dimensional impurity we showed numericaly the
possibility of excitation by the passing kink of non-linear
high-amplitude waves of new type, called here breathing pulson and
breathing 2D soliton. We suggested analytical expressions to model
these new excitations. The breathing pulson and breathing 2D
soliton radiate extended waves and their amplitudes slowly
decrease. Both excitations are long-living and can be of both
symmetric and asymmetric type depending on the impurity type. We
determined the range of the impurity parameters where the
breathing pulson and breathing 2D soliton can be excited. The
amplitude, size and frequency of the excited localized nonlinear
as the functions of the impurity parameters were given.

\section*{Acknowledgements}
\label{Acknowledgements}

The work was partly supported by the Russian Foundation for Basic
Research, grants 11-08-97057-p-povolzhie-a and 10-02-00594-a.

%% References


\bibliographystyle{model1-num-names}
\bibliography{bibliography}






% \end{thebibliography}

\end{document}